\documentclass[12pt]{article}
\usepackage{amssymb,slashed,latexsym,amsmath}
\usepackage[vcentermath,enableskew]{youngtab}
\newcommand{\ft}[2]{{\textstyle\frac{#1}{#2}}}
\def\Re{\mathop{\rm Re}\nolimits}
\def\Im{\mathop{\rm Im}\nolimits}
\def\trace{\mathop{\rm Tr}\nolimits}
\def\rme{{\rm e}}
\def\rmi{{\rm i}}
\def\rmd{{\rm d}}
\newsavebox{\uuunit}
\sbox{\uuunit}
    {\setlength{\unitlength}{0.825em}
     \begin{picture}(0.6,0.7)
        \thinlines
        \put(0,0){\line(1,0){0.5}}
        \put(0.15,0){\line(0,1){0.7}}
        \put(0.35,0){\line(0,1){0.8}}
       \multiput(0.3,0.8)(-0.04,-0.02){12}{\rule{0.5pt}{0.5pt}}
     \end {picture}}
\newcommand {\unity}{\mathord{\!\usebox{\uuunit}}}
\csname @addtoreset\endcsname{equation}{section}
\newcommand{\SO}{\mathop{\rm SO}}
\newcommand{\U}{\mathop{\rm {}U}}

   \usepackage{cite}
\begin{document}

 %%%%%%%%%%%%%%%%%%%%%%%%%%%%%%%%%%%%%%%%%%%%%%%%%%%%%%%%%%%
\begin{titlepage}
\begin{flushright}
KUL-TF-07/11\\
MPP-2007-61\\
arXiv:0705.4216
\end{flushright}
\vspace{.5cm}
\begin{center}
\baselineskip=16pt {\LARGE    Symplectic structure of $\mathcal{N}=1$ supergravity  \\
\vskip 0.2cm with anomalies and Chern-Simons terms
}\\
\vfill%\vskip 15mm%27.mm
{\large Jan De Rydt $^1$, Jan Rosseel $^1$, Torsten T. Schmidt $^{2}$, \\
\vskip 0.2cm
 Antoine Van Proeyen $^1$ and Marco Zagermann $^{2}$ %\\ \vskip 0.2cm between lines
  } \\
\vfill%\vskip 7mm%1cm
{\small $^1$ Instituut voor Theoretische Fysica, Katholieke Universiteit Leuven,\\
       Celestijnenlaan 200D B-3001 Leuven, Belgium.
      \\[2mm]
      $^2$ Max-Planck-Institut f{\"u}r Physik, F{\"o}hringer Ring 6,  \\ 80805 M{\"u}nchen, Germany
\\ \vspace{6pt}
 }
\end{center}
\vfill
\begin{center}
{\bf Abstract}
\end{center}
{\small The general actions of matter-coupled $\mathcal{N}=1$
supergravity have Peccei-Quinn terms that may violate gauge and
supersymmetry invariance. In addition, $\mathcal{N}=1$ supergravity with
vector multiplets may also contain generalized Chern-Simons terms. These
have often been neglected in the literature despite their importance for
gauge and supersymmetry invariance. We clarify the interplay of
Peccei-Quinn terms, generalized Chern-Simons terms and quantum anomalies
in the context of $\mathcal{N}=1$ supergravity and exhibit conditions
that have to be satisfied for their mutual consistency.  This extension
of the previously known $\mathcal{N}=1$ matter-coupled supergravity
actions follows naturally from the embedding of the gauge group into the
group of symplectic duality transformations. Our results regarding this
extension provide the supersymmetric framework for studies of string
compactifications with axionic shift symmetries, generalized Chern-Simons
terms and quantum anomalies.} \vfill

\hrule width 3.cm \vspace{2mm}{\footnotesize \noindent e-mails:
\{Jan.DeRydt, Jan.Rosseel, Antoine.VanProeyen\}@fys.kuleuven.be,
\\\phantom{e-mails: }\{schto, zagerman\}@mppmu.mpg.de }
\end{titlepage}
\addtocounter{page}{1}
 \tableofcontents{}
\newpage
%%%%%%%%%%%%%%%%
\section{Introduction}

Matter couplings in low energy effective actions of string
compactifications generically depend on scalar fields, such as the
moduli. An important example of such a scalar field dependence is
provided by non-minimal kinetic terms for gauge fields\footnote{We use
the non-Abelian field strength $
 {\cal F}_{\mu \nu }^A = F_{\mu \nu }^A + W_\mu ^B W_\nu ^Cf_{BC}{}^A$, where $F_{\mu \nu }^A=
2\partial _{[\mu }W_{\nu ]}^A$ is the Abelian part. The tilde denotes the
Hodge dual, as is further specified in the appendix.} (here enumerated by
an index $A$),
\begin{equation}
  e^{-1}{\cal L}_1  =
   -\ft14\Re f_{AB} {\cal F}_{\mu \nu }^A {\cal F}^{\mu \nu \,B} +\ft14\rmi\Im f_{AB} {\cal F}_{\mu \nu }^A \tilde {\cal F}^{\mu \nu
   \,B}\,,
 \label{L1f}
\end{equation}
where the gauge kinetic function $f_{AB}(z)$ is a nontrivial function of
the scalar fields, $z^{i}$, which, in  $\mathcal{N}=1$ supersymmetry, has
to be holomorphic. The second term in (\ref{L1f}) is often referred to as
the Peccei-Quinn term.

If, under a gauge transformation with gauge parameter $\Lambda^{A}(x)$,
some of the $z^{i}$ transform nontrivially, this may induce a
corresponding gauge transformation of $f_{AB}(z)$. If this transformation
is of the form of a symmetric product of two adjoint representations of
the gauge group,
 \begin{equation}
\delta(\Lambda)  f_{AB}= \Lambda ^C\delta _C f_{AB}\,,\qquad \delta _C
f_{AB}=f_{CA}{}^D f_{BD} + f_{CB}{}^Df_{AD}\, , \label{conditionfinvC1}
\end{equation}
with $f_{CA}{}^{B}$ the structure constants of the gauge group,  the
kinetic term (\ref{L1f}) is obviously gauge invariant. This is what was
assumed in the action of general matter-coupled supergravity in
\cite{Cremmer:1983en}\footnote{This construction of general
matter-couplings has been reviewed in \cite{Kallosh:2000ve}. There, the
possibility (\ref{conditionfinvCb}) was already mentioned, but the extra
terms necessary for its consistency were not considered.}.

If one takes into account also other terms in the (quantum) effective
action, however, a more general transformation rule for $f_{AB}(z)$ may
be allowed:
\begin{equation}
\delta _C f_{AB}=\rmi C_{AB,C} +f_{CA}{}^D f_{BD} + f_{CB}{}^Df_{AD}\,.
\label{conditionfinvCb}
\end{equation}
Here, $C_{AB,C}$ is a constant real tensor symmetric in the first two
indices, which we will recognize  as a natural generalization in the
context of symplectic duality transformations.

If $C_{AB,C}$ is non-zero, this leads to a non-gauge invariance of the
Peccei-Quinn term in  ${\cal L}_1$:
\begin{equation}
\delta(\Lambda ) e^{-1}{\cal L}_1  = \ft 14 \rmi C_{A B,C} \Lambda ^C
 {\cal F}_{\mu \nu }^A \tilde {\cal F}^{\mu \nu \,B }\,.
 \label{delLambdaSf1}
\end{equation}
For rigid parameters, $\Lambda^{A}=\mathrm{const.}$, this is just a total
derivative, but for local gauge parameters, $\Lambda^{A}(x)$, it is
obviously not. If (\ref{L1f}) is part of a supersymmetric action,   the
gauge non-invariance (\ref{delLambdaSf1}) also induces a non-invariance
of the action under supersymmetry, as we will recall  in section
\ref{ss:reviewTensorc}.

In order to  understand how this broken gauge and supersymmetry
invariance can be restored, it is convenient to split the coefficients
$C_{AB,C}$  into a sum,
\begin{equation}
C_{AB,C}=C^{\rm(s)}_{AB,C}+C_{AB,C}^{\rm(m)}\,,\qquad C^{\rm(s)}_{AB,C}=
C_{(AB,C)}\,,\qquad C_{(AB,C)}^{\rm(m)}=0\,,
 \label{eq:constr1_c}
\end{equation}
where $C_{AB,C}^{\rm(s)}$ is completely symmetric and $C_{AB,C}^{\rm(m)}$
denotes the part of mixed symmetry \footnote{This corresponds to the
decomposition $%\[
\Yboxdim8pt\yng(2)\otimes
\Yboxdim8pt\yng(1)=\Yboxdim8pt\yng(3)\oplus \Yboxdim8pt\yng(2,1)$.%\,.\]
}.
Terms of the form (\ref{delLambdaSf1}) may then in principle be cancelled
by the following two mechanisms, or a combination thereof:
\begin{enumerate}
\item As was first realized in a similar context in $\mathcal{N}=2$ supergravity
in \cite{deWit:1985px} (see also the systematic analysis
\cite{deWit:1987ph}), the gauge variation due to a non-vanishing mixed
part, $C_{AB,C}^{\rm(m)}\neq 0$, may be cancelled by adding a generalized
Chern-Simons term (GCS term) that contains a cubic and a quartic part in
the vector fields
 \begin{equation}
  \mathcal{L}_{\rm CS}= \ft12 C^{\rm (CS)}_{AB,C}\varepsilon ^{\mu \nu \rho \sigma }
  \left(\ft13 W_\mu ^CW_\nu ^A F_{\rho \sigma }^B +\ft14f_{DE}{}^A W_\mu ^DW_\nu ^E W_\rho ^CW_\sigma
  ^B\right)\,.
 \label{SCS1}
\end{equation}
This term depends on a constant tensor $C^{\rm (CS)}_{AB,C}$, which has
also a mixed symmetry structure. The cancellation occurs provided the
tensors $C_{AB,C}^{\rm(m)}$ and $C_{AB,C}^{\rm(CS)}$ are the same. It has
been shown in \cite{Andrianopoli:2004sv} that such a term exists as well
in rigid $\mathcal{N}=1$ supersymmetry.
\item If the chiral fermion spectrum is anomalous under the gauge group,
the anomalous triangle diagrams lead to a non-gauge invariance of the
quantum effective action of the form
$d_{ABC}\Lambda^{C}\mathcal{F}^{A}_{\mu\nu} {\tilde{\mathcal{F}}}^{\mu\nu
B}$ with a symmetric\footnote{More precisely, the anomalies have a scheme
dependence. As reviewed in \cite{Anastasopoulos:2006cz} one can choose a
scheme in which the anomaly is proportional to $d_{ABC}$. Choosing a
different scheme is equivalent to the choice of another GCS term (see
item (i).). We will always work with a renormalization scheme in which
the quantum anomaly is indeed proportional to $d_{ABC}$.} tensor
%,
$d_{ABC}\propto \trace ( \{T_{A},T_{B} \} T_{C})$. If $C_{AB,C}^{\rm
(s)}= d_{ABC}$, this quantum anomaly cancels the symmetric part of
(\ref{delLambdaSf1}). This is the Green-Schwarz mechanism.
\end{enumerate}
As has recently been emphasized in \cite{Anastasopoulos:2006cz}, both the
Green-Schwarz mechanism and the GCS terms are generically needed to
cancel the anomalies in orientifold models with intersecting D-branes.
Moreover, it is argued in \cite{Anastasopoulos:2006cz} that non-vanishing
GCS terms might have observable consequences for certain variants of
$Z^{\prime}$ bosons. On the other hand, as described in
\cite{Andrianopoli:2004sv}, GCS terms may also arise in certain flux and
generalized Scherk-Schwarz compactifications. Finally, they also play a
role in the manifestly symplectic formulation of gauged supergravity with
electric and magnetic potentials and tensor fields introduced in
\cite{deWit:2005ub}.

In view of these applications, it is surprising that the full interplay
between gauge invariance and (local) supersymmetry in the presence of GCS
terms and quantum anomalies is still only partially understood. In fact,
before the work of \cite{Andrianopoli:2004sv}, supersymmetric  GCS terms
were only studied in the context of \emph{extended} supersymmetry
\cite{deWit:1985px,deWit:2002vt,deWit:2003hr,Schon:2006kz,Derendinger:2007xp,deWit:2007mt}.
We would like to point out, however, that  there is an important
qualitative difference between $\mathcal{N}=1$ and $\mathcal{N}\geq 2$
supersymmetry. In extended supersymmetry, the $C_{AB,C}$ of
(\ref{conditionfinvCb}) have no symmetric part. This was already pointed
out in \cite{deWit:1985px} for the vector multiplets in $\mathcal{N}=2$
supergravity, at least in the presence of a prepotential. The equation
$C_{AB,C}^{\rm (s)}=0$ is also the basis of the manifestly symplectic
formulation \cite{deWit:2005ub}, where it is motivated by constraints
known from $\mathcal{N}=8$ supergravity. In $\mathcal{N}=1$ supergravity,
by contrast,  we find that the symmetric part of $C_{AB,C}$ may  be
present and could in principle cancel quantum anomalies. This is
consistent with the above-mentioned results on  extended supergravity
theories, because only $\mathcal{N}=1$ supergravity has the chiral
fermions that could possibly produce these quantum anomalies.

It is the purpose of this paper to give a systematic discussion of the
structure of general $\mathcal{N}=1$ supersymmetry with anomaly
cancellation and GCS terms. We will do this for a general gauge kinetic
function and an arbitrary gauge group with quantum anomalies. We also
consider the full coupling to supergravity and discuss its embedding into
the framework of the symplectic duality transformations. This generalizes
the work of \cite{Andrianopoli:2004sv}, which was restricted to linear
gauge kinetic functions of theories without quantum anomalies and to
rigid supersymmetry. As far as supersymmetry is concerned, the quantum
anomalies of the gauge symmetries are as important  as a classical
violation of gauge invariance, because the quantum anomalies of the gauge
symmetries also lead to supersymmetry anomalies as a consequence of the
supersymmetry algebra. The consistent gauge and supersymmetry anomalies
have been found for supergravity in \cite{Brandt:1993vd}. Our result for
the non-invariance of the sum of the kinetic terms and GCS terms in the
classical action matches with the results of \cite{Brandt:1993vd}.
%Brandt's formula.
\bigskip

The organization of the paper is as follows. In section
\ref{ss:symplectic} we explain how symplectic transformations act in
$\mathcal{N}=1$ supersymmetry, and how this leads to the generalized
transformation (\ref{conditionfinvCb}) of the gauge kinetic function
$f_{AB}$.

In the subsequent three sections, we first consider rigid supersymmetry.
More concretely, in section \ref{ss:reviewTensorc} we explore the
non-invariance of the kinetic terms of the vector multiplets under gauge
and supersymmetry transformations caused by (\ref{conditionfinvCb}). In
section \ref{ss:CS}, the GCS action and its role in the restauration of
gauge and supersymmetry invariance are discussed. Thirdly, in section
\ref{s:anomalies}, we consider the quantum anomaly as obtained in
\cite{Brandt:1993vd,Brandt:1997au}. Finally, we analyse the complete
cancellation of the gauge and supersymmetry anomalies by using the
results of the two previous sections.

The generalization to supergravity is considered in section
\ref{ss:sugracorr}. It turns out that the GCS terms obtained before can
just be added to the general actions of matter-coupled supergravity.

To show how this works in practice, it is useful to look at a gauge group
that is the product of an Abelian and a semisimple group. This setup was
also considered in \cite{Anastasopoulos:2006cz,Anastasopoulos:2007qm} and
\cite{Freedman:2005up,Elvang:2006jk}. Our discussion in section
\ref{ss:AbelSSimple} is close to the last reference, where it is
mentioned that local counterterms turn the consistent mixed anomalies
into a covariant mixed anomaly. This is the form of the anomaly that
appears as variation of the vector multiplet kinetic terms. The GCS terms
that we consider are precisely the counterterms that are mentioned in
\cite{Elvang:2006jk}.

We finish with conclusions and remarks in section \ref{ss:remarks} and
some notational issues are summarized in the appendix.

%%%%%%%%%%%%%%%%%%%%%%%%%%%%%%%%%%%%%%%%%%%%%%%%%%%%%%%%%%%%%

%%%%%%%%%%%%%%%%%%%%%%%%%%%%%%%%%%%%%%%%%%%%%%%%%%%%%%%%%%%%%

\section{Symplectic transformations in $\mathcal{N}=1$ supersymmetry}
\label{ss:symplectic}

In this section, we derive the general  form (\ref{conditionfinvCb}) of
the gauge transformation of the gauge kinetic function from the viewpoint
of symplectic duality  transformations. We begin by recalling the
essential elements of the duality transformations in four
dimensions~\cite{Ferrara:1977iq,deWit:1979sh,Cremmer:1979up,Gaillard:1981rj}.
The general form of kinetic terms for vector fields can be written in
several ways\footnote{The duality transformations, and hence the formulae
in the first part of this section, apply to the ungauged action.}:
\begin{eqnarray}
 e^{-1}{\cal L}_1 & = & -\ft14\Re f_{AB} F_{\mu \nu }^A F^{\mu \nu \,B} +\ft14\rmi\Im f_{AB} F_{\mu \nu }^A \tilde F^{\mu \nu \,B}   \nonumber\\
   & = & -\ft12 \Re \left(f_{AB}F_{\mu \nu }^{-\,A} F^{\mu \nu\,-\,B}\right)
   =-\ft12 \Im \left(F_{\mu \nu }^{-\,A} G^{\mu \nu\,-}_A\right)\,,
 \label{L1}
\end{eqnarray}
where the dual field strength is defined as
\begin{equation}
  G_A^{\mu \nu \,-}=-2\rmi\frac{\partial e^{-1}{\cal L}_1}{\partial F_{\mu \nu
  }^{-\,A}}=\rmi f_{AB}F^{\mu \nu  \,-\,B}\,.
 \label{defG}
\end{equation}
This shows that the Bianchi identities and field equations can be written
as
\begin{eqnarray}
\partial^\mu \Im {F}^{A\,- }_{\mu\nu} &=&0\ \ \ \ \ {\rm Bianchi\
identities,}\nonumber\\
\partial_\mu \Im G_{A }^{\mu\nu\,-} &=&0\ \ \ \ \  {\rm Equations\  of\
motion.}\label{BianchiField}
\end{eqnarray}
The set (\ref{BianchiField}) is invariant under the duality
transformations
\begin{equation}
\begin{pmatrix}F^{\prime -}\cr G^{\prime -}\end{pmatrix}%\ \rightarrow \
={\mathcal S} \begin{pmatrix}{F}^{-}\cr G^{-}
\end{pmatrix}=
\begin{pmatrix}A&B\cr C&D\end{pmatrix} \begin{pmatrix}{F}^-\cr
G^-\end{pmatrix}\,,\label{FGsympl}
\end{equation}
where the real matrices $A$, $B$, $C$ and $D$ satisfy
\begin{equation}
A^T C-C^T A=0\,,\qquad  B^T D- D^T B=0\,,\qquad   A^T D-C^T B=\unity\,.
\label{symplABCD}
\end{equation}
This guarantees that ${\cal S}$ is a symplectic matrix. In order to have
$G^{\prime}$ of the form (\ref{defG}), the kinetic matrix $f_{AB}$ is
transformed into $f'_{AB}$, where
\begin{equation}
\label{eq:sympl_f} \rmi f' = (C+D\rmi f)(A+B\rmi f)^{-1}\,.
\end{equation}

Symmetries of the action (\ref{L1}) correspond to symplectic
transformations with $B=0$, for which the Lagrangian (\ref{L1})
transforms into itself plus a total derivative if $C\neq 0$:
\begin{eqnarray}
e^{-1}{\mathcal{L}}^{\prime}_{1} &=&-\ft{1}{2} \Im (F_{\mu\nu}^{\prime - A} G_{A}^{\prime \mu\nu -})\nonumber \\
 & = & -\ft{1}{2} \Im (F_{\mu\nu}^{-A}G_{A}^{\mu\nu-}+ F_{\mu\nu}^{-A}(C^{T}A)_{AB}F^{B\mu\nu-})\,. \label{transformedL}
\end{eqnarray}
Not all of these rigid symmetries of the action can be promoted to gauge
symmetries. For this to be possible, the field strengths $F_{\mu\nu}^{A}$
have to transform in the adjoint representation of the prospective gauge
group. This determines the upper line of the transformation
(\ref{FGsympl}). We do not know a priori the transformation rule of
$f_{AB}$ and hence of $G_{\mu \nu \,A}$. The conditions
(\ref{symplABCD}), however, restrict further the corresponding symplectic
matrices to a form, which, at the infinitesimal level, reads
\begin{equation}
  {\cal S}=\unity -\Lambda ^C{\cal S}_C\,,\qquad {\cal S}_C=\begin{pmatrix}f_{CB}{}^A&0\cr
  C_{AB,C}&-f_{CA}{}^B\end{pmatrix}\,,
 \label{SC}
\end{equation}
where $C_{AB,C}$ is a real undetermined tensor, symmetric in its first
two indices. According to (\ref{eq:sympl_f}), the kinetic matrix should
then transform under the gauge transformations as
\begin{equation}
\delta %_{\mathrm{G}}
(\Lambda)  f_{AB}= \Lambda ^C\delta _C f_{AB}\,,\qquad
\delta _C f_{AB}=\rmi C_{AB,C} +f_{CA}{}^D f_{BD} + f_{CB}{}^Df_{AD}\,.
\label{conditionfinvC}
\end{equation}
The last two terms state that $f_{AB}$ transforms in the symmetric
product of two adjoint representations. The first term is the correction
to this and corresponds to the possible generalization by axionic shift
symmetries mentioned in the introduction. Note that the gauge kinetic
function might now transform nontrivially also under Abelian symmetries.

The algebra of gauge transformations is
\begin{equation}
  \left[ \delta (\Lambda _1),\delta (\Lambda _2)\right] =\delta
  (\Lambda_3^C=\Lambda _2^B\Lambda _1^A f_{AB}{}^C)\,.
 \label{algebratransf}
\end{equation}
In order that this algebra is realized by the symplectic transformations
(\ref{SC}), the commutators of the matrices ${\cal S}_A$ should be of the
form
\begin{equation}
  \left[ {\cal S}_A,{\cal S}_B\right] =f_{AB}{}^C{\cal S}_C\,.
 \label{SScomm}
\end{equation}
Written in full, this includes the equation
\begin{equation}
C_{AB,E}f_{CD}{}^E -2  C_{AE,[C}f_{D]B}{}^E -2  C_{BE,[C}f_{D]A}{}^E= 0
\,,
 \label{NonAbelianCident}
\end{equation}
which is the consistency condition that can be obtained by acting with
$\delta _D$ on (\ref{conditionfinvC}) and antisymmetrizing in $[CD]$.

Whether or not  the $C_{AB,C}$ can really be non-zero in a gauge theory,
and to what extent this  could be consistent with $\mathcal{N}=1$
supersymmetry is the subject of the remainder of this paper.

We finally note that, in this section, we considered only the vector
kinetic terms. The symplectic formulation gives also insight into other
terms of the action, which has been explored in \cite{Ceresole:1995jg}.
The additional terms to the action that we will discuss in this paper do
not modify this analysis. This is due to the fact that these new terms do
not involve the auxiliary fields $D$, while the analysis of
\cite{Ceresole:1995jg} is essentially dependent on the terms that result
from the elimination of these auxiliary fields.

%%%%%%%%%%%%%%%%%%%%%%%%%%%%%%%%%%%%%%%%%%%%%%%%%%%%%%%%%%

%%%%%%%%%%%%%%%%%%%%%%%%%%%%%%%%%%%%%%%%%%%%%%%%%%%%%%%%%%

\section{Kinetic terms of the vector multiplet}
 \label{ss:reviewTensorc}

Allowing for a nonvanishing shift $\,\,\rmi C_{AB,C}$ in
$\,\,\delta_{C}f_{AB}$ breaks both the gauge and supersymmetry
invariance. In this section, we make this statement more precise and
begin our discussion with some subtleties associated with the superspace
formulation in the Wess-Zumino gauge.

\subsection{The action}

The vector multiplet in the $\mathcal{N}=1$ superspace formulation is
described by a real superfield. The latter has many more components than
the physical fields describing an on-shell vector multiplet, which
consists of one vector field and one fermion. The advantage of this
redundancy is that one can easily construct manifestly supersymmetric
actions as integrals over full or chiral superspace. As an example
consider the expression
\begin{equation}
  S_f =\int \rmd^4 x \rmd^2 \theta\, f_{AB}(X) W_\alpha ^AW_\beta ^B \varepsilon
  ^{\alpha \beta } +\ c.c.
 \label{Sf}
\end{equation}
Here, $W_\alpha ^A= \ft14\bar D^2D_\alpha V^A$, or a generalization
thereof for the non-Abelian case, where $V^A$ is the real superfield
describing the vector multiplets labelled by an index $A$. The $f_{AB}$
are arbitrary holomorphic functions of a set of chiral superfields
denoted by $X$.

The integrand of (\ref{Sf}) is itself a chiral superfield. As we
integrate over a chiral superspace, the Lagrangian transforms into a
total derivative under supersymmetry. Formally, this conclusion holds
independently of the gauge symmetry properties of the functions
$f_{AB}(X)$. For the action (\ref{Sf}) to be gauge invariant, we should
have the condition \cite{Cremmer:1983en}
\begin{equation}
  \delta _C f_{AB}- f_{CA}{}^D f_{DB}- f_{AD}f_{CB}{}^D =0\,,
 \label{conditionfinv}
\end{equation}
where $\delta_C$ denotes the gauge transformation under the gauge
symmetry related to the vector multiplet denoted by the index $C$ as in
(\ref{conditionfinvC}).

Due to the large number of fields in the superspace formulation, the
gauge parameters are not just real numbers, but are themselves full
chiral superfields. To describe the physical theory, one wants to get rid
of these extra gauge transformations and thereby also of many spurious
components of the vector superfields. This is done by going to the
so-called Wess-Zumino gauge \cite{Wess:1974jb}, in which these extra
gauge transformations are fixed and many spurious components of the real
superfields are eliminated. Unfortunately, the Wess-Zumino gauge also
breaks the manifest supersymmetry of the superspace formalism. However, a
combination of this original ``superspace supersymmetry'' and the gauge
symmetries survives and becomes the preserved supersymmetry after the
gauge fixing. The law that gives the preserved supersymmetry as a
combination of these different symmetries is called the `decomposition
law', see e.g. eq. (2.28) in \cite{Cremmer:1983en}. Notice, however, that
this preservation requires the gauge invariance of the original action
(\ref{Sf}). Thus, though (\ref{Sf}) was invariant under the superspace
supersymmetry for any choice of $f_{AB}$, we now need
(\ref{conditionfinv}) for this action to be invariant under supersymmetry
after the Wess-Zumino gauge.

This important consequence of the Wess-Zumino gauge can also be
understood from the supersymmetry algebra. The superspace operator
$Q_\alpha $ satisfies the anticommutation relation
\begin{equation}
  \left\{Q_\alpha , Q^\dagger_{\dot \alpha } \right\}= \sigma_{\alpha \dot \alpha
  }^\mu \partial _\mu \,.
 \label{QQdagger}
\end{equation}
This equation shows no mixing between supersymmetry and gauge symmetries.
However, after the Wess-Zumino gauge the right-hand side is changed to
\cite{deWit:1975nq}
\begin{equation}
  \left\{Q_\alpha , Q^\dagger_{\dot \alpha } \right\}= \sigma_{\alpha \dot \alpha
  }^\mu {\cal D} _\mu = \sigma_{\alpha \dot \alpha
  }^\mu \left( \partial  _\mu -W_\mu ^A \delta _A\right)  \,,
 \label{QQdaggerD}
\end{equation}
where $\delta _A$ denotes the gauge transformation. Equation
(\ref{QQdaggerD}) implies that if an action is invariant under
supersymmetry, it should also be gauge invariant. \bigskip
\bigskip

As mentioned  before, the preservation of the Wess-Zumino gauges implies
that the effective supersymmetry transformations are different from the
ones in the original superspace formulation. It is shown in
\cite{deWit:1975nq} that the resulting supersymmetry transformations of a
chiral multiplet are
\begin{eqnarray}
 \delta (\epsilon ) z^i & = & \bar \epsilon _L\chi^i _L\,, \nonumber\\
 \delta (\epsilon ) \chi^i _L & = & \ft12\gamma ^\mu\epsilon _R {\cal D}_\mu
 z^i
 + \ft12 h^i\epsilon _L\,,\nonumber\\
 \delta (\epsilon ) h^i &=& \bar \epsilon _R\slashed{\cal D}
 \chi^i _L +
  \bar \epsilon _R \lambda _R^A \delta _A z^i\,,
 \label{susyChiralMult}
\end{eqnarray}
where we have denoted the scalar fields of the chiral multiplets as
$z^i$, the left-chiral components of the corresponding fermions as $\chi
_L^i$ and the auxiliary fields as $h^i$, while $\lambda ^A$ is the
gaugino of the vector multiplet $V^A$. These transformations are valid
for \textit{any} chiral multiplet, in particular, they can be applied to
the full integrand of (\ref{Sf}) itself. We will make use of this in
section \ref{ss:susytransfo}.

Compared to the standard superspace transformations, there are two
modifications in (\ref{susyChiralMult}). The first modification is that
the derivatives of $z^{i}$ and $\chi^{i}_{L}$ are covariantized with
respect to gauge transformations. This covariant derivative acts on   the
chiral fermions $\chi^i _L$  as
\begin{equation}
  {\cal D}_\mu \chi^i _L = \partial _\mu \chi^i _L - W_\mu ^A \delta _A\chi^i_L\,.
 \label{DmuOmega}
\end{equation}
Here, the gauge variation of the chiral fermions, $\delta _A\chi^i_L$,
can be expressed in terms of the gauge variation, $\delta_{A} z^{i}$, of
the scalar fields, using the fact that supersymmetry and gauge
transformations commute,
\begin{equation}
 \delta (\epsilon )\delta _A z^i  = \delta _A \delta (\epsilon )z^i =\delta _A\bar \epsilon _L\chi^i _L=
\bar \epsilon _L\delta _A\chi^i _L\,.
 \label{commdelAsusy}
\end{equation}
This leads to
\begin{equation}
  \delta _A \chi ^i = \frac{\partial \delta _A z^i}{\partial z^j}\chi
  ^j\,.
 \label{delAchi}
\end{equation}

The second modification is the additional last term in the transformation
of the auxiliary fields $h^i$. The origin of this term lies in the
contribution of the decomposition law for one of the gauge symmetries
contained in the chiral superfield of transformations $\Lambda $, after
the Wess-Zumino gauge is fixed.

To avoid the above-mentioned subtleties associated with the Wess-Zumino
gauge, we will use component field expressions in the remainder of this
text. Therefore, we reconsider the action (\ref{Sf}) and in particular
its integrand. The components of this composite chiral multiplet are
\cite{Cremmer:1983en}
\begin{eqnarray}
 z(fW^2) & = & -\ft12 f_{AB}\bar \lambda ^A_L\lambda ^B_L\,, \nonumber\\
 \chi_L(fW^2)  & = & \ft12 f_{AB}\left( \ft12 \gamma ^{\mu \nu }{\cal F}_{\mu \nu
 }^{A} -\rmi D^A\right) \lambda _L^B -\ft12 \partial_i f_{AB}\chi ^i_L \bar\lambda ^A_L\lambda ^B_L\,, \nonumber\\
 h(fW^2)&=&f_{AB}\left(-\bar \lambda ^A_L\slashed{\cal D}\lambda _R^B -\ft12 {\cal F}_{\mu \nu }^{-A}
 {\cal F}^{\mu \nu \,-B}+\ft12 D^AD^B\right) +\partial_i f_{AB}\chi ^i_L \left( -\ft12 \gamma ^{\mu \nu }{\cal F}_{\mu \nu
 }^{A} +\rmi D^A\right)\lambda ^B_L \nonumber\\
 && -\ft12 \partial_i f_{AB}h^i\bar\lambda ^A_L\lambda ^B_L +\ft12 \partial^2_{ij}
 f_{AB}\bar \chi _L^i\chi _L^j\bar\lambda ^A_L\lambda ^B_L \,,
 \label{componentszf}
\end{eqnarray}
where we used the notation $\partial_i=\frac{\partial}{\partial z^i}$.
The superspace integral in (\ref{Sf}) means that the real part of
$h(fW^2)$ is (proportional to) the Lagrangian:
\begin{equation}
  S_f= \int \rmd^4 x\, \Re h(fW^2)\,.
 \label{Sfh}
\end{equation}
{}From (\ref{componentszf}) and (\ref{Sfh}), we read off the kinetic
terms of $S_{f}$:
\begin{eqnarray}
S_{f,{\rm kin}}&=& \int \rmd^4x \Big[ -\ft14 \Re f_{AB}
\mathcal{F}_{\mu\nu}^{A}\mathcal{F}^{\mu\nu B} -\ft12 \Re f_{AB}
{\bar{\lambda}}^{A}
\slashed{\mathcal{D}}\lambda^{B}\nonumber   \\
&&\phantom{\int \rmd^4x \Big[}+ \ft14 \rmi \Im f_{AB}
\mathcal{F}_{\mu\nu}^{A}{\tilde{\mathcal{F}}}^{\mu\nu B} + \ft14 \rmi
(\mathcal{D}_{\mu} \Im f_{AB}) {\bar{\lambda}}^{A}\gamma^{5} \gamma^{\mu}
\lambda^{B} \Big]\,. \label{Sfkin}
\end{eqnarray}
In comparison to \cite{Cremmer:1983en}, we have used a partial
integration to shift the derivative from the gaugini to $(\Im f_{AB})$
and rearranged the structure constants in the last term, so as to obtain
a
``covariant'' derivative acting on $(\Im f_{AB})$. More precisely, we define
\begin{equation}
\label{derivkinfunction} {\cal D}_\mu f_{AB} = \partial_\mu f_{AB} - 2
W_\mu^C f_{C(A}{}^D f_{B)D}\,.
\end{equation}
In the case that the gauge kinetic matrix  transforms without a shift, as
in  (\ref{conditionfinv}), the derivative defined in
(\ref{derivkinfunction}) is fully gauge covariant.
\bigskip

In section \ref{ss:symplectic}, we motivated a more general gauge
transformation rule for $f_{AB}$, in which axionic shifts proportional to
$C_{AB,C}$ are allowed\footnote{We should remark here that
\cite{Andrianopoli:2004sv} restrict their work to the case in which
$f_{AB}$ is at most linear in scalars, and these scalars undergo a shift.
This is the most relevant way in which (\ref{conditionfinvC}) can be
realized.} as in (\ref{conditionfinvC}). Then (\ref{derivkinfunction}) is
no longer the full covariant derivative. The full covariant derivative
has instead the new form
\begin{equation}
 \hat{\cal D}_\mu f_{AB} \equiv \partial_{\mu} f_{AB} -W_{\mu}^{C}\delta_{C} f_{AB}
 = {\cal D}_\mu f_{AB} -  \rmi W_{\mu}^{C} C_{AB,C}\,.
 \label{fullcovderf}
\end{equation}
The last term in (\ref{Sfkin}) is therefore not gauge covariant for
non-vanishing $C_{AB,C}$. Hence, in presence of the new term in the
transformation of $f_{AB}$ we replace the action $S_f$ with $\hat{S}_f$,
in which we use the full covariant derivative, $\hat{\cal D}_\mu $,
instead of ${\cal D}_\mu $. More precisely, we define
\begin{equation}
  \hat{S}_f= S_f+S_{\rm{extra}}\,, \qquad S_{\rm extra}=\int \rmd^4 x\left(-\ft14 \rmi W_\mu^C
C_{AB,C} \bar\lambda^A \gamma_5\gamma^\mu \lambda^B\right)\,.
 \label{hatSf}
\end{equation}
Note that we did not use any superspace expression to derive
$S_{\rm{extra}}$ but simply added $S_{\rm{extra}}$ by hand in order to
fully covariantize the last term of (\ref{Sfkin}). As we will further
discuss in the next section,  $S_{\rm{extra}}$ can in fact only be
partially understood from superspace expressions, which motivates our
procedure to introduce it here by hand. We should also stress that the
covariantization with $S_{\rm{extra}}$ does not yet mean that the entire
action $\hat{S}_{f}$ is now fully gauge invariant. The gauge and
supersymmetry transformations of $\hat{S}_{f}$ will be discussed in
section~\ref{ss:susytransfo}.
\bigskip

We would finally like to emphasize that, in the context of
$\mathcal{N}=1$ supersymmetry, there is a priori no further restriction
on the symmetry of $C_{AB,C}$ apart from its symmetry in the first two
indices. This, however, is different in extended supersymmetry, as is
most easily demonstrated for $\mathcal{N}=2$ supersymmetry, where the
gauge kinetic matrix depends on the complex scalars $X^A$ of the vector
multiplets. These transform themselves in the adjoint representation,
which implies
\begin{equation}
  \delta (\Lambda )f_{AB}(X)= X^E \Lambda ^C f_{EC}{}^D\partial
  _Df_{AB}(X)\,.
 \label{delLambdafN2}
\end{equation}
Hence, this gives, from (\ref{conditionfinvC}),
\begin{equation}
  \rmi C_{AB,C} =X^Ef_{EC}{}^D\partial
  _Df_{AB}(X)-f_{CA}{}^D f_{BD} - f_{CB}{}^Df_{AD}\,,
 \label{CN2}
\end{equation}
which leads to $C_{AB,C}X^AX^BX^C=0$. As the scalars $X^A$ are
independent in rigid supersymmetry\footnote{The same argument can be made
for supergravity in the symplectic bases in which there is a
prepotential. However, that is not the case in all symplectic bases.
Bases that allow a prepotential are those were $X^A$ can be considered as
independent \cite{Ceresole:1995jg,Craps:1997gp}. An analogous argument
for other symplectic bases is missing. This is remarkable in view of the
fact that spontaneous breaking to $\mathcal{N}=1$ needs a symplectic
basis that allows no prepotential \cite{Ferrara:1995gu}. Hence, for the
$\mathcal{N}=2$ models that allow such a breaking to the $\mathcal{N}=1$
theories that we are considering in this paper, there is also no similar
argument for the absence of a totally symmetric part in $C_{AB,C}$,
except that for $\mathcal{N}=2$ there are no anomalies that could cancel
the corresponding gauge variation, due to the non-chiral nature of the
interactions.}, this implies that $C_{(AB,C)}=0$.

\subsection{Gauge and supersymmetry transformations}
\label{ss:susytransfo}

The action $S_f$ is gauge invariant before the modification of the
transformation of $f_{AB}$. In the presence of the $C_{AB,C}$ terms, the
action $\hat{S}_f$ is not gauge invariant. However, the non-invariance
comes only from one term. Indeed, terms in $\hat{S}_{f}$ that are
proportional to derivatives of $f_{AB}$ do not feel the constant shift
$\delta_{C} f_{AB} = \rmi C_{AB,C} + \ldots$. They are therefore
automatically gauge invariant. Also, the full covariant derivative
(\ref{fullcovderf}) has no gauge transformation proportional to
$C_{AB,C}$, and also $\Re f_{AB}$ is invariant. Hence, the gauge
non-invariance originates only from the third term in (\ref{Sfkin}). We
are thus left with
\begin{equation}
\delta(\Lambda ) \hat S_f  = \ft 14 \rmi C_{A B,C}\int \rmd^4x\, \Lambda
^C
 {\cal F}_{\mu \nu }^A \tilde {\cal F}^{\mu \nu \,B }\,.\label{delLambdaSf}
\end{equation}
This expression vanishes for constant $\Lambda$, but it spoils the
\emph{local} gauge invariance.
\bigskip

We started to construct $S_f$ as a superspace integral, and as such it
would automatically be supersymmetric. However, we saw that when $f_{AB}$
transforms with a shift as in (\ref{conditionfinvC}), the gauge symmetry
is broken, which is then communicated to the supersymmetry
transformations by the Wess-Zumino gauge fixing. The $C_{AB,C}$ tensors
then express the non-invariance of $S_f$ under both gauge transformations
and supersymmetry.

To determine these supersymmetry transformations, we consider the last
line of (\ref{susyChiralMult}) for $\{z^i,\chi^i, h^i\}$ replaced by
$\{z(fW^2), \chi (fW^2),h(fW^2)\}$ and  find
 \begin{equation}
  \delta(\epsilon ) S_{f}= \int \rmd^4 x \Re\left[
   \bar \epsilon_R \slashed{\partial}\chi_L(fW^2)-\bar \epsilon_R \gamma ^\mu W_\mu ^A \delta _A\chi_L(fW^2)
  +  \bar \epsilon _R \lambda _R^A \delta _A z(fW^2)\right] \,.
 \label{delepsSf}
\end{equation}
The first term in the transformation of $h(fW^2)$ is the one that was
already present in the superspace supersymmetry before going to
Wess-Zumino gauge. It is a total derivative, as we would expect from the
superspace rules. The other two terms are due to the mixing of
supersymmetry with gauge symmetries. They vanish if $z(fW^2)$ is
invariant under the gauge symmetry, as this implies by
(\ref{commdelAsusy}) that $\chi (fW^2)$ is also gauge invariant.

Using (\ref{componentszf}) and (\ref{conditionfinvC}), however,    one
sees that $z(fW^2)$ is not gauge invariant, and  (\ref{delepsSf})
becomes, using also (\ref{delAchi}),
\begin{eqnarray}
  \delta(\epsilon ) S_{f}&=& \int \rmd^4 x \Re \Big{\{} \rmi C_{AB,C}\Big{[}
   -\bar \epsilon_R \gamma ^\mu W_\mu ^C  \left( \ft14 \gamma ^{\rho \sigma}{\cal F}_{\rho \sigma
 }^{A} -\ft12\rmi D^A\right) \lambda _L^B %-\ft12 \partial_i C_{AB,C}\chi ^i_L \bar\lambda ^A_L\lambda ^B_L\Big)\nonumber\\
%&&\phantom{\int \rmd^4 x \Im}
-\ft12  \bar \epsilon _R \lambda _R^C \bar \lambda^A_L\lambda ^B_L\Big{]}
\Big{\}} \,.
 \label{delepsSfresult}
\end{eqnarray}
Note that this expression contains only fields of the vector multiplets
and none of the chiral multiplets.

It remains to determine the contribution of $S_{extra}$ to the
supersymmetry variation, which turns out to be
\begin{equation}
\delta(\epsilon)S_{\rm{extra}}=\int \rmd^4 x \Re \rmi C_{AB,C}\Big{[}
 -\ft12  W_\mu^C
\bar\lambda^B_L \gamma^\mu \left(\ft12 \gamma^{\nu
\rho}\mathcal{F}^A_{\nu \rho}- \rmi D^A\right)\epsilon_R-\bar\epsilon_R
\lambda_R^B \bar  \lambda_L^C \lambda^A_L\Big{]}\,.
\end{equation}
By combining this with (\ref{delepsSfresult}),  we obtain, after some
reordering,
\begin{equation}
\delta(\epsilon) \hat S_f  =\int \rmd^4 x \Re \left( \ft12
C_{AB,C}\varepsilon ^{\mu \nu \rho \sigma } W_\mu ^C {\cal F}_{\nu
\rho}^{A}\bar\epsilon_R \gamma _\sigma \lambda _L^B-\ft32 \rmi C_{(AB,C)}
\bar \epsilon_R \lambda _R^C \bar \lambda^A_L\lambda ^B_L \right) \,.
\label{deltaL}
\end{equation}
In sections \ref{ss:CS} and \ref{s:anomalies},  we describe how the
addition of GCS terms and quantum anomalies  can cancel the left-over
gauge and supersymmetry non-invariances of equations (\ref{delLambdaSf})
and (\ref{deltaL}).

%%%%%%%%%%%%%%%%%%%%%%%%%%%%%%%%%%%%%%%%%%%%%%%%%%%%%%%%%%%%%%%%%%%

%%%%%%%%%%%%%%%%%%%%%%%%%%%%%%%%%%%%%%%%%%%%%%%%%%%%%%%%%%%%%%%%%%%

\section{Chern-Simons action}
 \label{ss:CS}
\subsection{The action}
 \label{ss:CSaction}
Due to the gauged shift symmetry of $f_{AB}$, terms proportional to
$C_{AB,C}$ remain in the gauge and supersymmetry variation of the action
$\hat{S}_f$. To re-establish the gauge symmetry and supersymmetry
invariance, we need two ingredients: GCS terms and quantum anomalies. The
former were in part already discussed in
\cite{deWit:1985px,deWit:1987ph,Andrianopoli:2004sv}. They are of the
form
\begin{equation}
  S_{\rm CS}= \int \rmd^4 x\, \ft12 C^{\rm (CS)}_{AB,C}\varepsilon ^{\mu \nu \rho \sigma }
  \left(\ft13 W_\mu ^CW_\nu ^A F_{\rho \sigma }^B +\ft14f_{DE}{}^A W_\mu ^DW_\nu ^E W_\rho ^CW_\sigma
  ^B\right)\,.
 \label{SCS}
\end{equation}
The GCS terms are proportional to a tensor $C_{AB,C}^{\rm (CS)}$ that is
symmetric in $(A,B)$. Note that a completely symmetric part in
$C_{AB,C}^{\rm(CS)}$ would drop out of $S_{\rm CS}$ and we can therefore
restrict $C_{AB,C}^{\rm(CS)}$ to be a tensor of mixed symmetry structure,
i.e. with
\begin{equation}
  C^{\rm (CS)}_{(AB,C)}= 0\,.
 \label{Csymm0}
\end{equation}

A priori, the constants $C_{AB,C}^{\rm (CS)}$ need not be the same as the
$C_{AB,C}$ introduced in the previous section. For $\mathcal{N}=2$
supergravity \cite{deWit:1985px} one needs them to be the same, but we
will, for $\mathcal{N}=1$, establish another relation between both, which
follows from supersymmetry and gauge invariance requirements.

As was described in  \cite{Andrianopoli:2004sv}, the GCS terms can be
obtained from a superfield expression:
\begin{eqnarray}
  &&S_{\rm CS}'=C_{AB,C}^{\rm (CS)}\int \rmd^4x\,\rmd^4\theta \, \left[-\ft23 V^C\Omega ^{AB}(V)
  +\left( f_{DE}{}^B V^C{\cal D}^\alpha V^A\bar {\cal D}^2\left( {\cal D}_\alpha
  V^D V^E\right)+\ c.c.\right)  \right] \,,\nonumber\\
  &&\Omega ^{AB}={\cal D}^\alpha V^{(A}W_\alpha ^{B)}+\bar {\cal D}_{\dot \alpha }
  V^{(A}\bar W^{\dot \alpha B)}+ V^{(A}{\cal D}^\alpha W_\alpha ^{B)}\,.
 \label{SCSsuperspace}
\end{eqnarray}

The full non-Abelian superspace expression (\ref{SCSsuperspace}) is valid
only in the Wess-Zumino gauge, where it reduces to the bosonic component
expression (\ref{SCS}) plus a fermionic term \cite{Andrianopoli:2004sv}:
\begin{equation}
S_{\rm CS}^{\prime} = S_{\rm CS}+\left(S_{\rm CS}'\right)_{\rm
ferm}\,,\qquad \left(S_{\rm CS}'\right)_{\rm ferm}=\int \rmd^4
x\left(-\ft14 \rmi C_{AB,C}^{\rm(CS)} W_\mu^C \bar\lambda^A
\gamma_5\gamma^\mu \lambda^B\right)\,, \label{SCferm}
\end{equation}
where we used the restriction $C_{(AB,C)}^{\rm(CS)}=0$ from
(\ref{Csymm0}).

Note that the fermionic term in (\ref{SCferm}) is of a form similar to
$S_{\rm{extra}}$ in (\ref{hatSf}). More precisely, in   (\ref{SCferm})
the fermions appear with the tensor $C_{AB,C}^{\rm(CS)}$, which has a
mixed symmetry, (\ref{Csymm0}). $S_{\rm{extra}}$ in  (\ref{hatSf}), on
the other hand, is proportional to the tensor $C_{AB,C}^{(s)} +
C_{AB,C}^{(m)}$. From this we see that if we identify $C_{AB,C}^{(m)} =
C_{AB,C}^{\rm(CS)}$, as we will do later, we can absorb the mixed part of
$S_{\rm{extra}}$ into the superspace expression $S_{\rm CS}^{\prime}$.
This is, however, not possible for the symmetric part of $S_{\rm{extra}}$
proportional to $C_{AB,C}^{(s)}$, which cannot be obtained in any obvious
way from a superspace expression. As we need this symmetric part later,
it is more convenient to keep the full $S_{\rm extra}$, as we did in
section \ref{ss:reviewTensorc}, as a part of $\hat{S}_f$, and not include
$\left(S_{\rm CS}'\right)_{\rm ferm}$ here. Thus, we will further work
with the purely bosonic $S_{\rm CS}$ and omit the fermionic term that is
included in the superspace expression (\ref{SCSsuperspace}).
\bigskip

As an aside, we will show in the remainder of this subsection that for
semisimple algebras the GCS terms do not bring anything new
\cite{deWit:1987ph}, at least in the classical theory. By this we mean
they can be replaced by a redefinition of the kinetic matrix $f_{AB}$.
This argument is not essential for the main result of this paper and the
reader can thus skip this part. It shows, however, that the main
application of GCS terms is for non-semisimple gauge algebras.

We start with the result \cite{deWit:1987ph} that if
\begin{equation}
  C^{\rm (CS)}_{AB,C}= 2 f_{C(A}{}^DZ_{B)D}\,,
 \label{CinS}
\end{equation}
for a constant real symmetric matrix $Z_{AB}$, the action $S_{\rm CS}$
can be reabsorbed in the original action $S_f$ using
\begin{equation}
  f'_{AB}=f_{AB}+\rmi Z_{AB}\,.
 \label{fprime}
\end{equation}
In fact, one easily checks that with the substitution (\ref{CinS}) in
(\ref{conditionfinvC}), the $C$-terms are absorbed by the redefinition
(\ref{fprime}). The equation (\ref{CinS}) can be written as
\begin{equation}
  C^{\rm (CS)}_{AB,C}= T_{C,AB}{}^{DE}Z_{DE}\,,\qquad T_{C,AB}{}^{DE}\equiv 2f_{C(A}{}^{(D}
  \delta _{B)}^{E)}\,.
 \label{CisTS}
\end{equation}
In the case that the algebra is \emph{semisimple}, one can always
construct a $Z_{AB}$ such that this equation is valid for any
$C_{AB,C}^{\rm (CS)}$:
\begin{equation}
  Z_{AB}=C_2(T)^{-1}_{AB}{}^{CD}T_{E,CD}{}^{GH}g^{EF}C_{GH,F}^{\rm (CS)}\,,
 \label{SinC}
\end{equation}
where $g^{AB}$ and $C_2(T)^{-1}$ are the inverses of
\begin{equation}
  g_{AB}= f_{AC}{}^D f_{BD}{}^C\,,\qquad C_2(T)_{CD}{}^{EF}= g^{AB}
  T_{A,CD}{}^{GH}T_{B,GH}{}^{EF}\,.
 \label{defgC2}
\end{equation}
These inverses exist for semisimple groups. To show that (\ref{SinC})
leads to (\ref{CisTS}) one needs (\ref{NonAbelianCident}), which leads to
\begin{equation}
  g^{HD}T_H\cdot \left( \ft12 C^{\rm (CS)}_C f_{DE}{}^C+ T_{[D}\cdot C^{\rm (CS)}_{E]}\right)
  =0\,,
 \label{conseq1NAbident}
\end{equation}
where we have dropped doublet symmetric indices using the notation $\cdot
$ for contractions of such double indices. This further implies
\begin{equation}
  g^{AB}T_E\cdot T_B\cdot C^{\rm (CS)}_A= C_2(T)\cdot C^{\rm (CS)}_E\,,
 \label{conseq2Nabident}
\end{equation}
with which the mentioned conclusions can easily be obtained.

\subsection{Gauge and supersymmetry transformations}
The GCS term $S_{\rm CS}$ is not gauge invariant. Even the superspace
expression $S_{\rm CS}^{\prime}$  is not gauge invariant, not even in the
Abelian case. So, just as for $S_{f}$, we expect that $S_{\rm
CS}^{\prime}$ is not supersymmetric in the Wess-Zumino gauge, despite the
fact that it is a superspace integral. This is highlighted, in
particular, by the second term in (\ref{SCSsuperspace}), which involves
the structure constants. Its component expression simply gives the
non-Abelian $W\wedge W \wedge W \wedge W$ correction in (\ref{SCS}),
which, as a purely bosonic object, cannot be supersymmetric by itself.

For the gauge variation of $S_{\rm CS}$, one obtains
 \begin{eqnarray}
\nonumber\lefteqn{\delta(\Lambda) S_{\rm CS}=}\\
&&\int \rmd^4 x
\Big{[}-\ft{1}{4}\rmi C_{AB,C}^{\rm (CS)}\Lambda^C
F_{\mu\nu}^A \tilde F^{\mu \nu B} \label{eq:varNA}\\
\nonumber &&\phantom{\int \rmd^4 x \Big{[}}-\ft{1}{8}\Lambda^C\Big{(}2
C_{AB,D}^{\rm (CS)} f_{CE}{}^B
-  C_{DA,B}^{\rm (CS)}f_{CE}{}^B  + C_{BE,D}^{\rm (CS)}f_{CA}{}^B- C_{BD,C}^{\rm (CS)}f_{AE}{}^B \\
\nonumber&&\phantom{\int \rmd^4 x \Big{[}
-\ft{1}{8}\rmi\Lambda^C\Big{(}}+ C_{BC,D}^{\rm (CS)}f_{AE}{}^B+
C_{AB,C}^{\rm (CS)}f_{DE}{}^B +\ft12 C_{AC,B}^{\rm
(CS)}f_{DE}{}^B\Big{)}\varepsilon^{\mu\nu\rho\sigma}F_{\mu\nu}^AW_\rho^DW_\sigma^E
\\&& \phantom{\int \rmd^4 x \Big{[}}-\ft{1}{8}
\Lambda^C \Big{(}C_{BG,F}^{\rm (CS)}f_{CA}{}^B + C_{AG,B}^{\rm
(CS)}f_{CF}{}^B %\\&&
 %\phantom{\int \rmd^4 x \Big{[}-\ft{1}{8}\rmi\Lambda^C \Big{(}}
  + C_{AB,F}^{\rm (CS)}f_{CG}{}^B\Big{)} f_{DE}{}^A
\varepsilon^{\mu\nu\rho\sigma} W_\mu^D W_\nu^E W_\rho^F
W_\sigma^G\Big{]}\,,\nonumber
\end{eqnarray}
where we used the Jacobi identity %,
%\begin{equation}
%  2f_{D[A}{}^E f_{B]C}{}^D = f_{CD}{}^E f_{AB}{}^D\,,
%\end{equation}
and the property $C_{(AB,C)}^{\rm (CS)}=0$.

A careful calculation finally shows that the supersymmetry variation of
$S_{\rm CS}$ is
\begin{equation}
\delta(\epsilon) S_{\rm CS}=-\ft12\int \rmd^4 x\,
\varepsilon^{\mu\nu\rho\sigma}\Re \left[  C_{AB,C}^{\rm (CS)}W_\mu^C
F_{\nu \rho }^A  + C_{A[B,C}^{\rm (CS)}f_{DE]}{}^A \, W^E_\mu W^C_\nu
W_\rho ^D \right] \bar\epsilon_L \gamma_\sigma \lambda_R^B
\label{deltaLCS}\,.
\end{equation}

%%%%%%%%%%%%%%%%%%%%%%%%%%%%%%%%%%%%%%%%%%%%%%%%%%%%%%%%%

%%%%%%%%%%%%%%%%%%%%%%%%%%%%%%%%%%%%%%%%%%%%%%%%%%%%%%%%%

\section{Anomalies and their cancellation}
\label{s:anomalies}

In this section, we combine the classical non-invariances of $(\hat
S_{f}+S_{\rm CS})$ with the non-invariances induced by quantum anomalies.

\subsection{The consistent anomaly}

The physical information of a quantum field theory is contained in the
Green's functions, which in turn are encoded in an appropriate generating
functional. Treating the Yang-Mills fields $W_{\mu}$ as external fields,
the generating functional (effective action)  for proper   vertices
 can be  written as a path integral over the other matter fields,
\begin{equation}
\rme^{-\Gamma[W_\mu]}=\int {\cal D}\bar\phi{\cal D}\phi \rme^{-{\cal
S}(W_\mu, \bar\phi, \phi)}\, .
\end{equation}
The gauge invariance,
\begin{equation}
\delta_{A}\Gamma[W_{\mu}]=0\,, \label{Ward}
\end{equation}
 of the effective action encodes the  Ward identities and
  is crucial for the renormalizability of the theory. Even if the classical action, $\mathcal{S}$, is gauge invariant, a non-invariance of the path integral measure may occur and violate (\ref{Ward}), leading to a quantum anomaly. Even though the functional $\Gamma[W_{\mu}]$ is in general neither a  local nor a  polynomial functional  of the $W_{\mu}$, the
quantum anomaly,
\begin{equation}
\delta(\Lambda)\Gamma[W]=-\int \rmd^4x\,\Lambda^A \left({\cal D}_\mu
\frac{\delta \Gamma[W]}{\delta W_\mu}\right)_A \equiv \int
\rmd^4x\,\Lambda^A {\cal A}_A\,,
\end{equation}
does have this property. More explicitly, for an arbitrary non-Abelian
gauge group, the consistent form of the anomaly ${\cal A}_A$ is given by
\begin{equation}
\label{anomcon}{\cal A}_A \sim
\varepsilon^{\mu\nu\rho\sigma}\trace\Big{(}T_A \partial_\mu \left(W_\nu
\partial_\rho W_\sigma + \ft12 W_\nu W_\rho W_\sigma\right)\Big{)}\,,
\end{equation}
where $W_\mu =W_\mu ^A T_A$, and $T_A$ denotes the generators in the
representation space of the chiral fermions. Similarly there are
supersymmetry anomalies, such that the final non-invariance of the
one-loop effective action is
\begin{equation}
 {\cal A}= \delta \Gamma (W)= \delta(\Lambda)\Gamma[W]+\delta (\epsilon
  )\Gamma[W]=\int
\rmd^4x\,\left( \Lambda^A {\cal A}_A+\bar \epsilon {\cal
A}_\epsilon\right)  \,.
 \label{deltaGammaW}
\end{equation}
This anomaly should satisfy the Wess-Zumino consistency conditions
\cite{Wess:1971yu}, which are the statement that these variations should
satisfy the symmetry algebra. E.g. for the gauge anomalies these are:
\begin{equation}
\delta(\Lambda_1)\left(\Lambda_2^A {\cal
A}_A\right)-\delta(\Lambda_2)\left(\Lambda_1^A {\cal A}_A\right)=
\Lambda_1^B \Lambda_2^C f_{BC}{}^A {\cal A}_A\,.
\end{equation}
If the effective action is non-invariant under gauge transformations,
then also its supersymmetry transformation is non-vanishing. As we
explained in section \ref{ss:reviewTensorc}, this can for example be seen
from the algebra (\ref{QQdaggerD}).

A full cohomological analysis of anomalies in supergravity was made by
Brandt in \cite{Brandt:1993vd,Brandt:1997au}. His result (see especially
(9.2) in \cite{Brandt:1997au}) is that the total anomaly should be of the
form\footnote{This result is true up to local counterterms. The latter
are equivalent to a redefinition of the $C^{\rm (CS)}_{AB,C}$. This is
the same as the scheme-dependence mentioned in
\cite{Anastasopoulos:2006cz}, which is also equivalent to a modification
of these GCS terms.} (\ref{deltaGammaW}) with
\begin{eqnarray}
  {\cal A}_C&=&-\ft14\rmi \left[  d_{ABC} F_{\mu\nu}^B
+\left( d_{ABD}f_{CE}{}^B +\ft32 d_{ABC}f_{DE}{}^B \right)
W_\mu^DW_\nu^E\right]
\tilde F^{\mu\nu A}\,,\label{gaugeanom}\\
\bar \epsilon {\cal A}_\epsilon &=&\Re \left[ \ft32 \rmi d_{ABC}
\bar\epsilon_R \lambda_R^C \bar \lambda^A_L \lambda_L^B+\rmi
d_{ABC}W_\nu^C \tilde
F^{\mu\nu A} \bar\epsilon_L \gamma_\mu \lambda^B_R\right.\nonumber\\
&&\phantom{\Re }\left.+\ft{3}{8} d_{ABC}f_{DE}{}^A
\varepsilon^{\mu\nu\rho\sigma} W_\mu^D W_\nu^E W_\sigma^C \bar\epsilon_L
\gamma_\rho\lambda^B_R\right] \,.\label{susyanom}
\end{eqnarray}
The coefficients $d_{ABC}$ form a totally symmetric tensor that is not
fixed by the consistency conditions. Comparison with (\ref{anomcon})
implies that they are of the form
\begin{equation}
\label{dtrace} d_{ABC}\sim \trace\left(\{T_A,T_B\}T_C\right)\,.
\end{equation}

\subsection{The cancellation}
 \label{ss:cancellation}
Since the anomaly $\mathcal{A}$ is a local polynomial in $W_{\mu}$, one
might envisage a cancellation of the quantum anomaly by the classically
non-gauge invariant terms in the action in the spirit of the
Green-Schwarz mechanism.

The sum of the variations of the kinetic terms, (\ref{delLambdaSf}) and
(\ref{deltaL}), and of the variations of the GCS term, (\ref{eq:varNA})
and (\ref{deltaLCS}), simplifies if we set
\begin{equation}
  C^{\rm (CS)}_{ABC}= C^{\rm (m)}_{ABC}= C_{ABC} - C^{(s)}_{ABC}\,,
 \label{CCSsubst}
\end{equation}
and then use the consistency condition (\ref{NonAbelianCident}) for the
tensor $C_{ABC}$. The result is
\begin{eqnarray}
\nonumber\lefteqn{ \delta(\Lambda) \left(\hat S_f + S_{\rm CS}\right)=}\\
&&\ft14 \rmi \int \rmd^4x\, \Lambda^C \Big{[}C_{AB,C}^{\rm (s)}
F_{\mu\nu}^B +\Big{(}C_{AB,D}^{\rm (s)}f_{CE}{}^B +\ft32 C_{AB,C}^{\rm
(s)}f_{DE}{}^B \Big{)}W_\mu^DW_\nu^E\Big{]} \tilde F^{\mu\nu A}\,, \nonumber\\
\nonumber\lefteqn{ \delta(\epsilon) \left( \hat S_f+S_{\rm CS}\right)=}\\
%\nonumber&&
&&\int \rmd^4x\, \Re \Big{[} -\ft32 \rmi C_{AB,C}^{\rm (s)}
\bar\epsilon_R \lambda_R^C \bar \lambda^A_L \lambda_L^B- \rmi
C_{AB,C}^{\rm (s)}W_\nu^C
\tilde F^{\mu\nu A} \bar\epsilon_L \gamma_\mu \lambda^B_R \nonumber\\
&&\phantom{\int \rmd^4x\, \Big{[}}-\ft{3}{8} C_{AB,C}^{\rm (s)}f_{DE}{}^A
\varepsilon^{\mu\nu\rho\sigma} W_\mu^D W_\nu^E W_\sigma^C \bar\epsilon_L
\gamma_\rho\lambda^B_R\Big{]}\,. \label{totalgaugevar}
\end{eqnarray}
The integrand  of these expressions cancel the gauge anomaly
(\ref{gaugeanom}) and supersymmetry anomaly (\ref{susyanom}) if we set
\begin{equation}
\label{relCd} C_{AB,C}^{\rm (s)} = d_{ABC}\,.
\end{equation}

Thus, if $C_{AB,C}^{\rm (m)}=C_{AB,C}^{\rm (CS)}$ and $C_{AB,C}^{\rm (s)}
= d_{ABC}$, both gauge and supersymmetry are unbroken, in particular
anomaly-free. Note that this does not mean that any anomaly proportional
to some $d_{ABC}$ can be cancelled by a $C_{AB,C}^{\rm (s)}$. A gauge
kinetic function with an appropriate gauge  transformation induced by
gauge transformations of scalar fields such that (\ref{relCd}) holds may
simply not exist. Our analysis only shows that \emph{if} (\ref{relCd})
holds,   and $C_{AB,C}^{\rm (m)}=C_{AB,C}^{\rm (CS)}$ is satisfied, the
theory is gauge and supersymmetry invariant.

%%%%%%%%%%%%%%%%%%%%%%%%%%%%%%%%%%%%%%%%%%%%%%%%%%

%%%%%%%%%%%%%%%%%%%%%%%%%%%%%%%%%%%%%%%%%%%%%%%%%%

\section{Supergravity corrections}
 \label{ss:sugracorr}

In this section, we generalize our treatment to the full $\mathcal{N}=1$,
$d=4$ supergravity theory. We check supersymmetry and gauge invariance of
the supergravity action and show that no extra GCS terms (besides those
already added in the rigid theory) have to be included to obtain
supersymmetry or gauge invariance.

The simplest way to go from rigid supersymmetry to supergravity makes use
of the superconformal tensor calculus
\cite{Ferrara:1977ij,Kaku:1978nz,Kaku:1978ea,VanProeyen:1983wk}. A
summary in this context is given in \cite{Kallosh:2000ve}. Compared to
the rigid theory, the additional fields reside in a Weyl multiplet, i.e.
the gauge multiplet of the superconformal algebra, and a compensating
multiplet. The Weyl multiplet contains the vierbein, the gravitino
$\psi_{\mu}$ and an auxiliary vector, which will not be important for us.
The compensating multiplet enlarges the set of chiral multiplets in the
theory by one. The full set of fields in the chiral multiplets is now
$(X^I,\,\Omega ^I,\,H^I)$, which denote complex scalars, fermions and
complex auxiliary fields, respectively. The physical chiral multiplets
$(z^i,\chi^i, h^i)$ form a subset of these such that $I$ runs over one
more value than $i$. As our final results depend only on the vector
multiplet, this addition will not be very important for us, and we do not
have to discuss how the physical ones are embedded in the full set of
chiral multiplets.

When going from rigid supersymmetry to supergravity, extra terms appear
in the action (\ref{Sfh}); they are proportional to the gravitino
$\psi_{\mu}$. The integrand of (\ref{Sfh}) is replaced by the so-called
density formula, which is rather simple due to the use of the
superconformal calculus \cite{Ferrara:1978jt}:
\begin{equation}
\label{Sfsugra} S_f = \int \rmd^4x\, e\,\mbox{Re}\left[ h(fW^2) +
\bar\psi_{\mu R} \gamma^\mu \chi_L(fW^2) +  \ft12 \bar\psi_{\mu
R}\gamma^{\mu\nu}\psi_{\nu R}z(fW^2)\right]\, ,
\end{equation}
where $e$ is the determinant of the vierbein. For completeness, we give
the component expression of (\ref{Sfsugra}). It can be found by plugging
in the relations (\ref{componentszf}), where we replace the fields of the
chiral multiplets with an index $i$ by the larger set indexed by $I$,
into the density formula (\ref{Sfsugra}). The result is
\begin{eqnarray}
\nonumber \lefteqn{\hat S_f=\int \rmd^4x\,e}\\&&\nonumber \Big{[}\Re f_{A
B}(X)\left( -\ft14 {\cal F}_{\mu \nu }^A {\cal F}^{\mu \nu \,B } -\ft12
\bar \lambda^A \gamma^\mu \hat{\cal D}_\mu\lambda^B +\ft12 D^A
D^B+\ft18\bar\psi_\mu\gamma^{\nu\rho}\left({\cal F}_{\nu\rho}^A+\hat{\cal
F}_{\nu\rho}^A\right)\gamma^\mu\lambda^B\right)
\\\nonumber&&
\phantom{\Big{[}}+ \ft 14 \rmi \Im f_{A B}(X){\cal F}_{\mu \nu }^A
\tilde{\cal F}^{\mu \nu B}+\ft14 \rmi  \left(\hat{\cal D}_\mu \Im
f_{AB}(X)\right)\bar \lambda ^A \gamma _5\gamma ^\mu \lambda ^B
\\\nonumber&&
\phantom{\Big{[}}+\Big{\{} \ft12  \partial_I f_{A B}(X)\left[\bar
\Omega_L^I\left( -\ft12\gamma ^{\mu \nu } \hat{\cal F}_{\mu \nu }^{A
}+\rmi D^A \right)\lambda _L^B-\ft12  \left( H^I +\bar\psi_{\mu
R}\gamma^\mu \Omega^I_L\right)\bar \lambda_L^A\lambda _L^B\right]
\\\label{fullsugraaction}&&
\phantom{\Big{[}+\Big{\{}} +\ft14  \partial_I
\partial_J f_{A B }(X)\,\bar \Omega_L^I\Omega_L^J \bar \lambda _L^A \lambda
_L^B + {\rm h.c.} \Big{\}}\Big{]}\,,
\end{eqnarray}
where the hat denotes full covariantization with respect to gauge and
local supersymmetry, e.g.
\begin{equation}
\label{Fhat}\hat {\cal F}_{\mu \nu}^A = {\cal F}_{\mu \nu}^A +
\bar\psi_{[\mu}\gamma_{\nu]}\lambda^A\,.
\end{equation}
Note that we use already the derivative $\hat{\cal D}_\mu \Im f_{AB}(X)$,
covariant with respect to the shift symmetries, as explained around
(\ref{fullcovderf}). Therefore, we denote this action as $\hat S_f$ as we
did for rigid supersymmetry.

The kinetic matrix $f_{AB}$ is now a function of the scalars $X^I$. We
thus have in the superconformal formulation
\begin{equation}
  \delta_C f_{AB}=\partial _If_{AB}\delta _C X^I= \rmi C_{AB,C}+\ldots \,.
 \label{CtermsI}
\end{equation}

Let us first consider the supersymmetry variation of
(\ref{fullsugraaction}). Compared with (\ref{deltaL}), the supersymmetry
variation of (\ref{fullsugraaction}) can only get extra contributions
that are proportional to the $C$-tensor. These extra contributions come
from the variation of $H^I$ and $\Omega^I$ in covariant objects that are
now also covariantized with respect to the supersymmetry transformations
and from the variation of $e$ and $\lambda^A$ in the gauge
covariantization of the $({\hat{{\cal D}}}_\mu \Im f_{AB})$-term. Let us
list in more detail the parts of the action that give these extra
contributions.

First there is a coupling of $\Omega^I$ with a gravitino and gaugini,
coming from \\$- \ft14 e \partial_I f_{AB} \bar \Omega^I_L \gamma^{\mu
\nu} \hat {\cal F}_{\mu \nu}^A \lambda^B_L$:
\begin{eqnarray}
\nonumber\lefteqn{S_1 = \int\rmd^4x\, e \Big{[}-\ft14 \partial_I f_{AB}
\bar \Omega^I_L \gamma^{\mu \nu}\lambda^B_L
\bar\psi_{[\mu}\gamma_{\nu]}\lambda^A+\mbox{h.c.}\Big{]}}\\&&\rightarrow
\delta(\epsilon) S_1 = \int\rmd^4x\, e \Big{[}-\ft18 \rmi C_{AB,C}
W_\rho^C \bar \lambda^B_L \gamma^{\mu \nu}\gamma^\rho \epsilon_R
\bar\psi_\mu \gamma_\nu\lambda^A+\hdots+\mbox{h.c.}\Big{]}.
\end{eqnarray}
We used the expression (\ref{Fhat}) for $\hat {\cal F}_{\mu\nu}^A$ and
(\ref{susyChiralMult}) where ${\cal D}_\mu X^I$ is now also covariantized
with respect to the supersymmetry transformations, i.e. $\hat{{\cal
D}}_\mu X^I$. There is another coupling between $\Omega^I$, a gravitino
and gaugini that we will treat separately:
\begin{eqnarray}
\nonumber\lefteqn{S_2 = \int\rmd^4x\, e \Big{[}\ft14 \partial_I f_{AB}
\bar \Omega^I_L \gamma^\mu\psi_{\mu R} \bar\lambda^A_L\lambda^B_L
+\mbox{h.c.}\Big{]}}\\&&\rightarrow \delta(\epsilon) S_2=\int\rmd^4x\, e
\Big{[}\ft18 \rmi C_{AB,C} W_\rho^C \bar\epsilon_R \gamma^\rho \gamma^\mu
\psi_{\mu R} \bar \lambda_L^A \lambda_L^B+\hdots+\mbox{h.c.}\Big{]}.
\end{eqnarray}
A third contribution comes from the variation of the auxiliary field
$H^I$ in $S_3$, where
\begin{equation}
S_3 = \int\rmd^4x\,e\, \Big{[}-\ft14  \partial_I f_{AB} H^I
\bar\lambda^A_L\lambda^B_L +\mbox{h.c.}\Big{]} \, .
\end{equation}
The variation is of the form
\begin{equation}
\delta_\epsilon H^I = \bar\epsilon_R \gamma^\mu \mathcal{D}_\mu \Omega
^I_L+\hdots=-\ft12 \bar\epsilon_R \gamma^\mu \gamma^\nu
\hat{\mathcal{D}}_\nu X^I \psi_{\mu R}+\hdots = \ft12 \delta_C X^I
W_\nu^C \bar\epsilon_R \gamma^\mu \gamma^\nu \psi_{\mu R}+\hdots\,.
\end{equation}
Therefore we obtain
\begin{eqnarray}
\nonumber\lefteqn{S_3 = \int\rmd^4x\,e\, \Big{[}-\ft14  \partial_I f_{AB}
H^I \bar\lambda^A_L\lambda^B_L
+\mbox{h.c.}\Big{]}}\\&&\rightarrow\delta(\epsilon) S_3 =
\int\rmd^4x\,e\, \Big{[}- \ft18\rmi  C_{AB,C} W_\nu^C \bar \epsilon_R
\gamma^\mu\gamma^\nu \psi_{\mu R}\bar \lambda_L^A
\lambda_L^B+\hdots+\mbox{h.c.}\Big{]}.
\end{eqnarray}
Finally, we need to consider the variation of the vierbein $e$ and the
gaugini in a part of the covariant derivative on $\Im f_{AB}$:
\begin{eqnarray}
\nonumber\lefteqn{S_4 = \int\rmd^4x\,e\, \Big{[}\ft14\rmi C_{AB,C}
W_\mu^C \bar\lambda^A \gamma^\mu \gamma_5
\lambda^B\Big{]}}\\\nonumber&&\rightarrow\delta(\epsilon) S_4=\int\rmd^4x
\,e\,\Big{[}-\ft14\rmi  C_{AB,C} W_\rho^C\Big{(} \bar\lambda_R^A
\gamma^\mu \lambda_L^B \bar\epsilon_R \gamma^\rho \psi_{\mu L}+\ft{1}{4}
\bar\epsilon_R\gamma^\rho \gamma^\mu\gamma^\nu \psi_{\nu L}
\bar\lambda_L^A \gamma_\mu \lambda^B_R\\
\nonumber&&\phantom{\rightarrow\delta(\epsilon) S_4=\int\rmd^4x
\Big{[}-\ft14\rmi  C_{AB,C} W_\rho^C\Big{(}}+\ft{1}{4}
\bar\epsilon_R\gamma^\rho\gamma^\mu\psi_{\mu R}
\bar\lambda_{L}^A\lambda_L^B\Big{)}\\&&\phantom{\rightarrow\delta(\epsilon)
S_4=\int\rmd^4x \,e\,\Big{[}} +\ft{1}{4}\rmi  C_{AB,C} W^{\mu C}
\bar\psi_{\mu R}\epsilon_R
\bar\lambda_{L}^A\lambda_L^B+\hdots+\mbox{h.c.}\Big{]}\,.
\end{eqnarray}
It requires some careful manipulations to obtain the given result for
$\delta(\epsilon) S_4$. One needs the variation of the determinant of the
vierbein, gamma matrix identities and Fierz relations.

In the end, we find that $\delta(\epsilon)\left(S_1 + S_2 + S_3 +
S_4\right)=0$. This means that all extra contributions that were not
present in the supersymmetry variation of the original supergravity
action vanish without the need of extra terms (e.g. generalizations of
the GCS terms). We should also remark here that the variation of the GCS
terms themselves is not influenced by the transition from rigid
supersymmetry to supergravity because it depends only on the vectors
$W_\mu ^A$, whose supersymmetry transformations have no gravitino
corrections in $\mathcal{N}=1$.

Let us check now the gauge invariance of terms proportional to the
gravitino. Neither terms involving the real part of the gauge kinetic
function, $\Re f_{AB}$, nor its derivatives violate the gauge invariance
of $\hat{S}_{f}$. The only contributions to gauge non-invariance come
from the pure imaginary parts, $\Im f_{AB}$, of the gauge kinetic
function. On the other hand, no extra $\Im f_{AB}$ terms appear when one
goes from rigid supersymmetry to supergravity and, hence, the gauge
variation of $\hat{S}_f$ does not contain any gravitini. This is
consistent with our earlier result that neither
$\delta(\epsilon)\hat{S}_f$ nor $S_{\rm CS}$ contain gravitini.

Consequently, the general $\mathcal{N}=1$ action contains just the extra
terms (\ref{SCS}), and we can add them to the original action in
\cite{Cremmer:1983en}.

%%%%%%%%%%%%%%%%%%%%%%%%%%%%%%%%%%%%%%%%%%%%%%%%%%%%%%%%%%%%%%%%%%%%%%%%

%%%%%%%%%%%%%%%%%%%%%%%%%%%%%%%%%%%%%%%%%%%%%%%%%%%%%%%%%%%%%%%%%%%%%%%%

\section{Specializing to Abelian $\times $ semisimple gauge groups}
 \label{ss:AbelSSimple}
We mentioned at the end of section \ref{ss:CSaction} that simple gauge
groups do not lead to non-trivial GCS terms. Therefore we consider now a
relevant case: the product of a (one-dimensional) Abelian factor and a
semisimple gauge group. This will allow us to clarify the relation
between our results and previous work, in particular
\cite{Freedman:2005up,Elvang:2006jk}. In these papers, the authors study
the structure of quantum consistency conditions of $\mathcal{N}=1$
supergravity. More precisely, they clarify the anomaly cancellation
conditions (required by the quantum consistency) for a $\U(1) \times G$
gauge group, where $G$ is semisimple. We introduce the notations
$F_{\mu\nu}$ and ${\cal G}^a_{\mu\nu}$ for the Abelian and semisimple
field strengths, respectively.

In this case, one can look at
``mixed'' anomalies, which are the ones proportional to $\trace
(QT_aT_b)$, where $Q$ is the $\U(1)$ charge operator and $T_a$ are the
generators of the semisimple algebra. Following \cite[section
2.2]{Elvang:2006jk}, one can add counterterms such that the mixed
anomalies proportional to $\Lambda^a$ cancel and one remains with those
that are of the form $\Lambda^0 \trace\left(Q{\cal G}_{\mu\nu}\tilde
{\cal G}^{\mu\nu}\right)$, where $\Lambda^0$ is the Abelian gauge
parameter. Schematically, it looks like
\begin{equation}\label{freedmanstrategy1}
\begin{array}{c|ccc}
\mbox{Anomalies:}&\Lambda^a {\cal A}^a_{\rm mixed\,con}&+&\Lambda^0 {\cal
A}^0_{\rm mixed\,con}\\\hline
\delta(\Lambda){\cal L}_{\rm ct}:&-\Lambda^a {\cal A}^a_{\rm mixed\,con}&-&\Lambda^0 {\cal A}^0_{\rm mixed\,con}\\
&&+&\Lambda^0 {\cal A}^0_{\rm mixed\,cov}\\\hline\hline
\mbox{sum:}&0&+&\Lambda^0 {\cal A}^0_{\rm mixed\,cov}
\end{array}
\end{equation}
where the subscripts `con' and `cov' denote the consistent and covariant
anomalies, respectively. The counterterms ${\cal L}_{\rm ct}$ have the
following form:
\begin{equation}
{\cal L}_{\rm ct}=\ft13 Z\varepsilon^{\mu\nu\rho\sigma} C_\mu
\trace\Big{[}Q\left(W_\nu \partial_\rho W_\sigma + \ft34 W_\nu W_\rho
W_\sigma \right)\Big{]}\,,\qquad Z=\frac{1}{4\pi ^2}\,,\label{counter}
\end{equation}
where $C_\mu$ and $W_\mu$ are the gauge fields for the Abelian and
semisimple gauge groups respectively. The expressions for the anomalies
are:
\begin{eqnarray}
\nonumber{\cal A}^a_{\rm mixed\,con}&=&-\ft13Z\varepsilon^{\mu\nu\rho\sigma}\trace\Big{[}T^a Q \partial_\mu\left(C_\nu \partial_\rho W_\sigma + \ft14 C_\nu W_\rho W_\sigma\right)\Big{]}\,,\\
\nonumber{\cal A}^0_{\rm mixed\,con}&=&-\ft16Z\varepsilon^{\mu\nu\rho\sigma}\trace\Big{[}Q \partial_\mu\left(W_\nu \partial_\rho W_\sigma + \ft12 W_\nu W_\rho W_\sigma\right)\Big{]}\,,\\
{\cal A}^0_{\rm
mixed\,cov}&=&-\ft18\varepsilon^{\mu\nu\rho\sigma}\trace\Big{[}Q{\cal
G}_{\mu\nu}{\cal G}_{\rho \sigma }\Big{]}\,. \label{AFreedman}
\end{eqnarray}
The remaining anomaly ${\cal A}^0_{\rm mixed\,cov}$ is typically
cancelled by the Green-Schwarz mechanism.
\bigskip

We will compare this now with our results for general non-Abelian gauge
groups, which we reduce to the case Abelian $\times$ semisimple. The
index $A$ is split into $0$ for the $\U(1)$ and $a$ for the semisimple
group generators. We expect the GCS terms (\ref{SCS}) to be equivalent to
the counterterms in \cite{Elvang:2006jk} and the role of the
Green-Schwarz mechanism is played by a $\U(1)$ variation of the kinetic
terms $f_{ab}$, hence by a $C$-tensor with non-trivial components
$C_{ab,0}$.

It follows from the consistency condition (\ref{NonAbelianCident}) that
\begin{equation}
C_{0a,0}=C_{00,a}=0\, \label{C00a}
\end{equation}
and the $C_{ab,0}$'s are proportional to the Cartan-Killing metric in
each simple factor. We write here
\begin{equation}
C_{ab,0}= Z \trace(QT_aT_b)\,, \label{mixedC}
\end{equation}
where $Z$ could be arbitrary, but our results will match the results of
\cite{Elvang:2006jk} for the value of $Z$ in (\ref{counter}).

We will not allow for off-diagonal elements of the gauge kinetic function
$f_{AB}$:
\begin{equation}
f_{0a}=0 \hspace{2mm}\Rightarrow \hspace{2mm} C_{0a,b}=0\,.
\end{equation}
There may be non-zero components $C_{00,0}$ and $C_{ab,c}$, but we shall
be concerned here only with the mixed ones, i.e.\ we have only
(\ref{mixedC}) different from zero.

If we reduce (\ref{delLambdaSf}) using (\ref{C00a}) and (\ref{mixedC}) we
get
\begin{eqnarray}
\left[\delta(\Lambda) \hat S_f\right]_{\rm mixed} &=&\int \rmd^4x
\Big{[}\ft18
 Z \Lambda^0 \varepsilon^{\mu\nu\rho\sigma}\trace\left(Q{\cal
G}_{\mu\nu}{\cal G}_{\rho \sigma}\right)\Big{]}\,. \label{deltaSfmixed}
\end{eqnarray}
Splitting (\ref{mixedC}) into a totally symmetric and mixed symmetry part
gives
\begin{eqnarray}
  &&C^{\rm (s)}_{ab,0}=C^{\rm (s)}_{0a,b}=\ft13C_{ab,0}=\ft13Z
  \trace(QT_aT_b)\,, \nonumber\\
 && C^{\rm (m)}_{ab,0}=\ft23C_{ab,0}=\ft23Z  \trace(QT_aT_b)\,, \qquad
C^{\rm (m)}_{0a,b}=-\ft13C_{ab,0}=-\ft13Z
  \trace(QT_aT_b)\,.
 \label{Csmexample}
\end{eqnarray}
We learned in section \ref{ss:cancellation} that for a final gauge and
supersymmetry invariant theory we have to take $C^{\rm CS}=C^{\rm (m)}$,
and hence the mixed part of the GCS action (\ref{SCS}) reads in this
case:
\begin{eqnarray}
\left[S_{\rm CS}\right]_{\rm mixed}&=&\int \mbox{d}^4x \,\Big{[}\ft13 Z
C_\mu \varepsilon^{\mu\nu\rho\sigma}\trace \left[ Q\left(W_\nu
\partial_\rho W_\sigma + \ft34 W_\nu W_\rho W_\sigma\right)\right] \Big{]}\,.
\label{SCSexample}
\end{eqnarray}
Finally, we reduce the consistent anomaly (\ref{gaugeanom}) using
$d_{ABC}=C^{\rm(s)}_{ABC}$. We find
\begin{eqnarray}
{\cal A}_0&=&- \ft{1}{6} Z\varepsilon^{\mu\nu\rho\sigma}\trace\left[
Q\partial _\mu \left(W_\nu \partial _\rho W_\sigma+\ft12
 W_\nu W_\rho W_\sigma\right)\right]\,,  \nonumber\\
{\cal A}_a&=&-\ft{1}{3}Z \Lambda^a
\varepsilon^{\mu\nu\rho\sigma}\trace\left[  T_aQ\partial _\mu\left (
C_\nu\partial _\rho W_\sigma+\ft14 C_\nu  W_\rho W_\sigma\right)\right]
\,, \label{deltaGSf+SCSred}
\end{eqnarray}
where $G_{\mu\nu}$ is the Abelian part of the gauge field ${\cal G}_{\mu
\nu}$.

We can make the following observations:
\begin{enumerate}
\item The mixed part of the GCS action
(\ref{SCSexample}) is indeed equal to the counterterms (\ref{counter}),
introduced in \cite{Elvang:2006jk}.
\item
The consistent anomalies (\ref{deltaGSf+SCSred}), for which we based our
formula on \cite{Brandt:1993vd,Brandt:1997au}, match those in the first
two lines of (\ref{AFreedman}). As we mentioned above, the counterterm
has modified the resulting anomaly to the covariant form in the last line
of (\ref{AFreedman}).
\item
We see that the variation of the kinetic term for the vector fields
(\ref{deltaSfmixed}) is able to cancel this mixed covariant anomaly (this
is the Green-Schwarz mechanism).
\end{enumerate}

Combining these remarks, our cancellation procedure can schematically be
presented as follows:
\begin{equation}
\begin{array}{c|ccc}
\mbox{Anomalies:}&\Lambda^a {\cal A}^a_{\rm mixed\,con}&+&\Lambda^0 {\cal
A}^0_{\rm mixed\,con}\\\hline
\delta(\Lambda){\cal L}_{\rm (CS)}:&-\Lambda^a {\cal A}^a_{\rm mixed\,con}&-&\Lambda^0 {\cal A}^0_{\rm mixed\,con}\\
&&+&\Lambda^0 {\cal A}^0_{\rm mixed\,cov}\\\hline\delta(\Lambda)\hat
S_f:&&-&\Lambda^0 {\cal A}^0_{\rm mixed\,cov}\\\hline\hline
\mbox{sum:}&0&+&0
\end{array}
\end{equation}

\section{Conclusions}
 \label{ss:remarks}
In this paper, we have studied the consistency conditions that ensure the
gauge and supersymmetry invariance of matter coupled $\mathcal{N}=1$
supergravity theories with Peccei-Quinn terms, generalized Chern-Simons
terms % here in the conclusions, I would not write GCS
and quantum anomalies. Each of these three ingredients defines a constant
three index tensor:
\begin{enumerate}
  \item The gauge non-invariance of the Peccei-Quinn terms is proportional to
a constant imaginary shift of the gauge kinetic function parameterized by
a tensor $C_{AB,C}$. This tensor in general splits into a completely
symmetric part and a part of mixed symmetry, $C_{AB,C}^{\rm(s)} +
C_{AB,C}^{\rm(m)}$.
 \item Generalized Chern-Simons terms are defined by a
tensor, $C_{AB,C}^{\rm (CS)}$, of mixed symmetry.
\item Quantum gauge
anomalies of chiral fermions are proportional to a tensor $d_{ABC}$,
which, in the appropriate regularization scheme, can be chosen to be
completely symmetric, $d_{ABC} \propto \trace(\{T_{A},T_{B}\} T_{C})$.
\end{enumerate}
We find the full quantum effective action to be gauge invariant and
supersymmetric if
\begin{equation}
C_{AB,C}=C_{AB,C}^{\rm(CS)} + d_{ABC}\,.
\end{equation}

The inclusion of the quantum anomalies encoded in a non-trivial tensor
$d_{ABC}$ is the key feature that distinguishes $\mathcal{N}=1$ theories
from theories with extended supersymmetry. Because of their possible
presence, the Peccei-Quinn shift tensor $C_{AB,C}$ can now have a
nontrivial symmetric part, $C_{AB,C}^{\rm (s)}$. In the context of
$\mathcal{N}=2$ supergravity, the absence of such a completely symmetric
part can be directly proven for theories for which there exists a
prepotential \cite{deWit:1985px}.

We performed our analysis first in rigid supersymmetry. Using
superconformal techniques, we could then show that only one cancellation
had to be checked to extend the results to supergravity. It turns out
that the Chern-Simons term does not need any gravitino corrections and
can thus be added as such to the matter-coupled supergravity actions. Our
paper provides thus an extension to the general framework of coupled
chiral and vector multiplets in $\mathcal{N}=1$ supergravity.\footnote{We
should emphasize that we only considered anomalies of gauge symmetries
that are gauged by elementary vector fields. The interplay with K{\"a}hler
anomalies in supergravity theories can be an involved subject
\cite{Freedman:2005up,Elvang:2006jk}, which we did not study.}

Our results are interesting for a number of rather different
applications. For example, in reference \cite{deWit:2005ub}, a general
set-up for treating gauged supergravities in a manifestly symplectic
framework was proposed. In that work the completely symmetric part of
what we call $C_{AB,C}$ was assumed to be zero, following the guideline
of extended supergravity theories. As we emphasized in this paper,
$\mathcal{N}=1$ supergravity theories might allow for a non-vanishing
$C_{AB,C}^{(s)}$, and hence  a possible extension of the setup of
\cite{deWit:2005ub}  in the presence of quantum anomalies. It might be
interesting to see whether such an extension really exists.

In \cite{Anastasopoulos:2006cz}, orientifold compactifications with
anomalous fermion spectra were studied, in which the chiral anomalies are
cancelled by a mixture of the Green-Schwarz mechanism and generalized
Chern-Simons terms. The analysis in \cite{Anastasopoulos:2006cz} was
mainly concerned with the gauge invariance of the bosonic part of the
action and revealed the generic presence of a completely symmetric and a
mixed part in $C_{AB,C}$ and the generic necessity of generalized
Chern-Simons terms. Our results show how such theories can be embedded
into the framework of $\mathcal{N}=1$ supergravity and supplements the
phenomenological discussions of \cite{Anastasopoulos:2006cz} by the
fermionic couplings in a supersymmetric setting.

The work of   \cite{Anastasopoulos:2006cz} raises the general question of
the possible higher-dimensional origins of GCS terms. In
\cite{Andrianopoli:2004sv}, certain flux and generalized Scherk-Schwarz
compactifications \cite{D'Auria:2003jk,Angelantonj:2003rq} are identified
as another means to generate such terms. In \cite{Gunaydin:2005bf}, it
was also shown that $\mathcal{N}=2$ supergravity theories with GCS terms
can be obtained by ordinary dimensional reduction of certain $5D$,
$\mathcal{N}=2$ supergravity theories with tensor multiplets
\cite{Gunaydin:1999zx,Bergshoeff:2004kh}. It would be interesting to
obtain a more complete picture of the possible origins of GCS-terms in
string theory and supergravity theories.

%%%%%%%%%%%%%%%%%%%%%%%%%%%%%%%%
\medskip
\section*{Acknowledgments.}

\noindent We are grateful to Massimo Bianchi, Claudio Caviezel, Bernard
de Wit, Sergio Ferrara, Dan Freedman, Elias Kiritsis, Simon K{\"o}rs, Henning
Samtleben and Kelly Stelle for useful discussions. We are also grateful
for the hospitality of the M.P.I. in M{\"u}nchen and of the K.U. Leuven
during various stages of this work. J.R. and A.V.P. thank the Galileo
Galilei Institute for Theoretical Physics for hospitality and the INFN
for partial support.

This work is supported in part by the European Community's Human
Potential Programme under contract MRTN-CT-2004-005104 `Constituents,
fundamental forces and symmetries of the universe'.
%%
%%more optional lines:
J.R. is Aspirant FWO-Vlaanderen.
The work of J.d.R., J.R. and A.V.P. is supported in part by the FWO -
Vlaanderen, project G.0235.05 and by the Federal Office for Scientific,
Technical and Cultural Affairs through the `Interuniversity Attraction
Poles Programme -- Belgian Science Policy' P6/11-P. The work of T.S. and
M.Z. is supported by the German Research Foundation (DFG) within the
Emmy-Noether Programme (Grant number  ZA 279/1-2).

\newpage
%%%%%%%%%%%%%%%%%%%%%%%%%%%
\appendix
\section{Notation}
We follow closely the notation of \cite{Kallosh:2000ve}, but use real
$\varepsilon _{0123}=1$. The (anti)selfdual field strengths are the same
as there, i.e.
\begin{equation}
  F^\pm _{\mu \nu }=\ft12\left( F_{\mu \nu }\pm \tilde F_{\mu \nu }\right)
  \,, \qquad \tilde F^{\mu \nu }=-\ft12 \rmi e^{-1}\varepsilon ^{\mu \nu \rho \sigma }
  F_{\rho \sigma }\,.
 \label{F+-tilde}
\end{equation}
One difference is that we use indices $A,B,\ldots $ for gauge indices,
such that $\alpha ,\beta $ can be used for 2-component spinors in
superspace expressions. Square brackets around indices like $[AB]$ denote
the antisymmetrization with total weight one, thus for two indices it
includes a factor $1/2$ for each combination.

For comparison with Wess and Bagger notations, the $\gamma ^m$ are
\begin{equation}
  \gamma ^m= \begin{pmatrix}0&\rmi \sigma ^m\cr \rmi\bar \sigma
  ^m&0\end{pmatrix}\,,
 \label{gammamBW}
\end{equation}
where these sigma matrices are $\sigma ^m_{\alpha \dot \beta }$ or
$\sigma ^{m\dot \alpha \beta }$. Spinors that we use are in their
2-component notation
\begin{equation}
  \chi=\begin{pmatrix}\chi _\alpha \cr \bar \chi ^{\dot \alpha
}\end{pmatrix}\,, \qquad \bar \chi =\begin{pmatrix}\chi ^\alpha &\bar
\chi _{\dot \alpha }\end{pmatrix}\,,
 \label{spinorsBW}
\end{equation}
where $\bar \chi =\chi ^T C= \rmi \chi ^\dagger \gamma ^0$. Further
translation is obtained by replacing in Wess-Bagger
\begin{eqnarray}
&&  S\rightarrow S\,,\qquad \chi \rightarrow \sqrt{2}\chi \,,\qquad
F\rightarrow F\,, \nonumber\\
&& \lambda \rightarrow -\lambda \,,\qquad D\rightarrow -D\,,\qquad W_\mu
\rightarrow W_\mu \,,\nonumber\\
&&\epsilon \rightarrow \ft12\epsilon \,.
 \label{translationBW}
\end{eqnarray}

%%%%%%%%%%%%%%%%%%%%%%%%%%%%%%%%%%%%%%%%%%%%%%%%%%%%%%%%
%\bibliography{refLectParis}
%%Included for WinEdt Gather Purpose (do not remove the comment line below:
%             %input "C:\localtexmf\bibtex\bib\refLectParis.bib"
%             %input "C:\Program Files\MiKTeX\texmf\bibtex\bib\refLectParis.bib"
%\bibliographystyle{toine}

\providecommand{\href}[2]{#2}\begingroup\raggedright\endgroup

\end{document}